\documentclass[]{spie}  

\usepackage[]{graphicx,amssymb,amsmath}

\title{Chirped-pulse oscillators: an impact of the dynamic gain saturation}

\author{Vladimir L. Kalashnikov
\skiplinehalf Institut f\"{u}r Photonik, TU Wien, Gusshausstr.
27/387, A-1040 Vienna, Austria}

\authorinfo{Further author information:\\V.L.Kalashnikov:
E-mail: kalashnikov@tuwien.ac.at, Telephone: +43 1 588 01 387 43}

\begin{document}
  \maketitle

\begin{abstract}
An effect of the dynamic gain saturation on chirped-pulse oscillator
was investigated. It was found, that the dynamic gain saturation
causes strong perturbations of the pulse front that destabilizes an
oscillator. As a result, the chirped-pulse exists only within some
limited range of dispersions and there is a limit of energy growth
for a given resonator period.
\end{abstract}



\section{INTRODUCTION}
\label{intro}  

Oscillators directly generating femtosecond pulses with energy
exceeding 100 nJ are of interest for numerous applications including
frequency conversion, metrology, micro-machining, etc. Nowadays
oscillators operating at MHz-repetition rates have reached the
micro-Joule pulse energy frontier without an extra-resonator
amplification. An ingenious idea allowing such an advance is the
pulse stretching, that suppresses the instabilities inherent to the
high-peak-power oscillators \cite{fujimoto}. Since the pulse is
soliton in the negative group-delay-dispersion (GDD) regime, its
stretching is linearly irreversible. The issue of pulse
compressibility can be solved by a chirped-pulse oscillator (CPO)
providing: i) sufficient pulse stretching (few picoseconds), ii)
broad spectrum (more than 100 nm), and iii) pulse compressibility
down to few tens of femtoseconds \cite{ap1}.

The remaining issue is that a high-energy oscillator (including CPO)
is stable only within a confined region of GDD and pulse energy. The
existing theory \cite{k1,k2} predicts that some minimum positive GDD
is required for the CPO stabilization against the CW-amplification
and this ``threshold'' GDD increases with the energy. It was found,
that the gain saturation is a key factor determining the oscillator
stability. However, the existing models do not take into account the
gain dynamics and, thereby, some substantial features of the regime
remain unexplored. For instance, it is unclear why the region of GDD
and energy, where a single stable pulse exists, is confined
\cite{confined}.

Here, for the first time to our knowledge, the theory of CPO taking
into account the gain dynamics will be presented. As it will be
shown, the gain dynamics provides an explanation for the confinement
of the pulse stability region. The model demonstrates an existence
of two main mechanisms of pulse destabilization: the satellite rise
in front of the pulse and the CW-amplification.

\section{CPO MODELING} \label{s1}

The CPO conception is based on the pulse energy ($E$) scaling with
the scaling of resonator period ($T_{cav}$): $E = P_{av} T_{cav}$
($P_{av}$ is the averaged power). Since the pulse peak power
($P(0)$) scales with the energy: $P(0) \approx E/T$ ($T$ is the
pulse width), one has to stretch the pulse (i.e., to increase $T$)
in order to reduce $P(0)$ and, thereby, suppress the instabilities
causing by the nonlinear effects inside a resonator (e.g., the
self-focusing inside an active medium). However, in the solitonic
regime (i.e., when the GDD is negative), such a stretching is
linearly irreversible. This means that the femtosecond high-energy
pulse can be produced by only nonlinear compression, which worsens
its characteristics (e.g., enhances the noise). In the CPO, the
picosecond pulse has an extremely wide spectrum due to large chirp.
As a result, the pulse is linearly compressible down to $\approx
1/\Delta$ ($\Delta$ is the pulse spectrum width). The experiments
demonstrate that such a regime is very stable and reproducible
\cite{ap1}.

The pulse stability in the CPO can be explained by the combined
action of two mechanisms: a pure phase mechanism and a
frequency-dissipative one. The first mechanism is the balance of
phase contributions from the pulse envelope $A(t)$: $\beta
{{\partial ^2 A\left( t \right)} \mathord{\left/
 {\vphantom {{\partial ^2 A\left( t \right)} {\partial t^2 }}} \right.
 \kern-\nulldelimiterspace} {\partial t^2 }}$ ($\beta$ is the GDD coefficient, $t$ is the local time ranging from $-T_{cav}/2$ to $T_{cav}/2$) and the
 time-dependent phase $\phi (t)$: $ - \beta A\left( t \right)\left[ {{{\partial \phi \left( t \right)} \mathord{\left/
 {\vphantom {{\partial \phi \left( t \right)} {\partial t}}} \right.
 \kern-\nulldelimiterspace} {\partial t}}} \right]^2$ \cite{oe}. Such
 a balance is possible if the pulse is chirped. However, a pure phase balance is not
 sufficient for pulse stabilization as the pulse spreads, i.e.
 some dissipative effects are required to form a quasi-soliton. The pulse lengthening due to GDD can be compensated
 by its shortening  owing to frequency filtering  if the pulse is
 chirped \cite{haus,proctor}. The chirp results in frequency deviation at the pulse
 front and tail, and the filter cuts off the high- and
 low-frequency wings of the pulse, thus the pulse shortens.

The CPO can be described on the basis of the distributed generalized
complex cubic-quintic nonlinear Ginzburg-Landau model
\cite{k1,k2,akhmediev}:

\begin{equation}\label{eq1}
\frac{{\partial A (z,t)}}{{\partial z}} = \left[ {-\sigma(z,t)
A(z,t) + \left( {\alpha  - \beta } \right)\frac{{\partial ^2
A(z,t)}}{{\partial t^2 }}} \right] + \left[ {\left( {\kappa  -
i\gamma } \right)P(z,t) - \kappa \zeta P(z,t)^2 } \right]A(z,t).
\end{equation}

\noindent Here $z$ is the propagation distance normalized to the
cavity length (i.e., the cavity round-trip number, in fact).
$P(z,t)=|A(z,t)|^2$ is the instant power, $\alpha$ is the square of
the inverse spectral filter bandwidth. The parameters $\gamma$ and
$\kappa$ are the self-phase and self-amplitude modulation
coefficients, respectively. Parameter $\zeta$ describes saturation
of the self-amplitude modulation. In a broadband solid-state
oscillator, for example, parameter $\alpha$ is the square of the
inverse gain band width multiplied by the saturated gain
coefficient. Parameters $\kappa$ and $\zeta$ are defined by the
Kerr-lens mode locking mechanism \cite{k1,k2}.

The $\sigma$-parameter is the spectrally independent saturable
net-loss coefficient, which depends on the instant pulse energy
$E\left(z, t \right) \equiv \int\limits_{ - T_{cav}/2 }^t {P\left(
{z,t'} \right)dt'}$. Below, such a dependence will be referred as
the dynamic gain saturation.

It was found \cite{k2}, that a  heavily-chirped pulse in CPO can be
described as the soliton-like solution of Eq. (\ref{eq1}) in the
limit of time-independent $\sigma$. In this case, the simplest
energy-dependence of the $\sigma$-parameter is
\begin{equation}\label{eq2}
\sigma \left( E \right) \approx \delta \left( {\frac{E} {{E^* }} -
1} \right),
\end{equation}

\noindent where
 $ \delta \equiv
E^* \left. {\frac{{d\sigma }} {{dE}}} \right|_{E = E^* }$, $ E
\equiv \int\limits_{ - T_{cav} /2}^{T_{cav} /2} {P\left( {t'}
\right)dt'} $ and $E^*\equiv P_{av} T_{cav}$. The spectrum of the
chirped soliton-like pulse has a Lorentz profile truncated at some
frequency $\pm \Delta/2$, where the central frequency $\omega=$0
corresponds to the CPO carrier frequency \cite{k2}. The chirped
soliton-like pulse is stable within
 the region of its existence, i.e. within the region of
 $\sigma>0$ \cite{k2}. Positivity of $\sigma$, i.e. stability
 against the CW-amplification (vacuum stability), can be provided by a certain
 minimum positive GDD growing with energy.

Nevertheless, the existing theory does not take into account the
time dependence of $\sigma$ provided by the dynamic gain saturation.
Let us consider some numerical estimations. The gain saturation
fluency for a Ti:sapphire oscillator is $E_s  \equiv {{h\nu }
\mathord{\left/
 {\vphantom {{h\nu } {\sigma _g }}} \right.
 \kern-\nulldelimiterspace} {\sigma _g }} \approx
$0.8 J/cm$^2$ ($\sigma_g$ is the emission cross-section, $\nu$ is
the emission frequency, $h$ is the Planck's constant). Then, the
gain variation due to gain saturation per one resonator round-trip
is $ {T \mathord{\left/
 {\vphantom {T {S E_s \gamma }}} \right.
 \kern-\nulldelimiterspace} {S E_s \gamma }} \approx
$0.2 ($T=$1 ps, the mode area $S$ equals to 130 $\mu$m$^2$, $E=$200
nJ; and the peak power is close to the stability limit, which is
defined by the self-phase modulation coefficient $\gamma=$4.5
MW$^{-1}$). That is the gain variation per one cavity round-trip is
not negligible unlike that in a low-energy femtosecond oscillator
operating in the solitonic regime.

Below, the extended theory of CPO taking into account the dynamic
gain saturation will be considered.

\section{ANALYTICAL THEORY OF CPO WITH THE DYNAMIC GAIN SATURATION}
\label{s2}

The cubic nonlinear limit of Eq. (\ref{eq1}) admits the chirped
soliton-like solution \cite{haus,k3}:
\begin{equation}\label{eq3}
A\left( t \right) = \sqrt {P\left( 0 \right)} \operatorname{sech}
\left[ {{{\left( {t - \upsilon z } \right)} \mathord{\left/
 {\vphantom {{\left( {t - \upsilon z} \right)} T}} \right.
 \kern-\nulldelimiterspace} T}} \right]^{1 + i\psi } \exp \left[ {i\omega \left( {t - \upsilon z } \right)} +i\phi z
 \right],
\end{equation}
which is frequency ($\omega$)- and phase ($\phi$)-shifted and
time-delayed ($\upsilon$) in the presence of the dynamic gain
saturation ($\psi$ is the dimensionless chirp):

\begin{equation}\label{eq4}
\sigma \left( {z,t} \right) = l - g\left( {\upsilon z} \right) +
\frac{{1}}{{E_s}} \int\limits_{\upsilon z}^t {P\left( {z,t'}
\right)} dt'.
\end{equation}

\noindent The pulse parameters are:

\begin{equation}\label{eq5}
\begin{gathered}
  \psi  = \frac{{3\left( {{\kappa  \mathord{\left/
 {\vphantom {\kappa  {\gamma  - {\beta  \mathord{\left/
 {\vphantom {\beta  \alpha }} \right.
 \kern-\nulldelimiterspace} \alpha }}}} \right.
 \kern-\nulldelimiterspace} {\gamma  - {\beta  \mathord{\left/
 {\vphantom {\beta  \alpha }} \right.
 \kern-\nulldelimiterspace} \alpha }}}} \right) - \sqrt {9\left( {{{\beta ^2 } \mathord{\left/
 {\vphantom {{\beta ^2 } {\alpha ^2 }}} \right.
 \kern-\nulldelimiterspace} {\alpha ^2 }} + {{\kappa ^2 } \mathord{\left/
 {\vphantom {{\kappa ^2 } {\gamma ^2 }}} \right.
 \kern-\nulldelimiterspace} {\gamma ^2 }}} \right) + 8\left( {1 + \left( {{{\kappa \beta } \mathord{\left/
 {\vphantom {{\kappa \beta } {\gamma \alpha }}} \right.
 \kern-\nulldelimiterspace} {\gamma \alpha }}} \right)^2 } \right) - 2{{\kappa \beta } \mathord{\left/
 {\vphantom {{\kappa \beta } {\gamma \alpha }}} \right.
 \kern-\nulldelimiterspace} {\gamma \alpha }}} }}
{{2\left( {1 + {{\beta \kappa } \mathord{\left/
 {\vphantom {{\beta \kappa } {\gamma \alpha }}} \right.
 \kern-\nulldelimiterspace} {\gamma \alpha }}} \right)}}, \hfill \\
  \upsilon  = {{\left( {{{3\alpha \psi } \mathord{\left/
 {\vphantom {{3\alpha \psi } {E_s }}} \right.
 \kern-\nulldelimiterspace} {E_s }} + 2\alpha \omega \psi \gamma  - {{\psi ^2 \beta } \mathord{\left/
 {\vphantom {{\psi ^2 \beta } {E_s }}} \right.
 \kern-\nulldelimiterspace} {E_s }} + {{2\beta } \mathord{\left/
 {\vphantom {{2\beta } {E_s }}} \right.
 \kern-\nulldelimiterspace} {E_s }} + 2\omega \beta \gamma } \right)} \mathord{\left/
 {\vphantom {{\left( {{{3\alpha \psi } \mathord{\left/
 {\vphantom {{3\alpha \psi } {E_s }}} \right.
 \kern-\nulldelimiterspace} {E_s }} + 2\alpha \omega \psi \gamma  - {{\psi ^2 \beta } \mathord{\left/
 {\vphantom {{\psi ^2 \beta } {E_s }}} \right.
 \kern-\nulldelimiterspace} {E_s }} + {{2\beta } \mathord{\left/
 {\vphantom {{2\beta } {E_s }}} \right.
 \kern-\nulldelimiterspace} {E_s }} + 2\omega \beta \gamma } \right)} \gamma }} \right.
 \kern-\nulldelimiterspace} \gamma }, \hfill \\
  \omega  = \frac{{\psi \left( {\beta \psi ^2  - 3\alpha \psi  - 2\beta } \right)}}
{{2\gamma \alpha E_s \left( {1 + \psi ^2 } \right)}},\,\,T^2  =
\frac{{\alpha \left( {1 - \psi ^2 } \right) - 2\beta \psi }}
{{\alpha \omega ^2  + \sigma_0 }},\,\,P\left( 0 \right) =
\frac{{\beta \left( {\psi ^2  - 2} \right) - 3\alpha \psi }}
{{\gamma T^2 }}. \hfill \\
\end{gathered}
\end{equation}

\noindent In Eq. (\ref{eq4}) $l$ is the net-loss coefficient,
$g(\upsilon z)$ is the gain at the pulse peak taking into account
its time-delay with propagation, $\sigma_0= l - g(\upsilon z)$. One
can see a clear manifestation of the dynamic gain saturation: the
time-delay and the frequency shift appear.

\begin{figure}
   \begin{center}
   \begin{tabular}{c}
   \includegraphics[height=6cm]{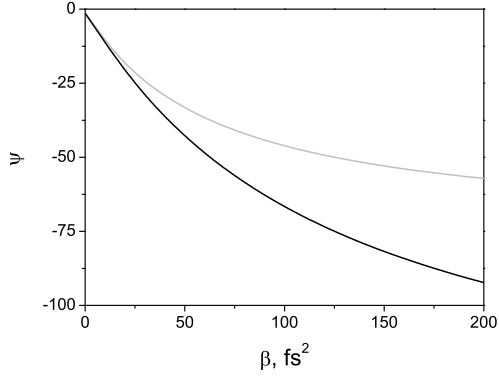}
   \end{tabular}
   \end{center}
   \caption[f1]
   { \label{f1}
Dependence of the chirp $\psi$ on GDD for the model
(\ref{eq1},\ref{eq3},\ref{eq4}) with $\zeta=$0 and $\alpha=$2.5
fs$^2$. $\kappa=$0.02$\gamma$ (black) and 0.04$\gamma$ (gray).}
   \end{figure}

Fig. \ref{f1} shows the GDD-dependence of the chirp parameter $\psi$
for two fixed $\kappa$. One can see, that the chirp ($|\psi|$)
increases with GDD and its value is sufficiently large to provide an
efficient compression (the pulse width after compression is $\approx
T/|\psi|$). Growth of the self-amplitude modulation decreases
$|\psi|$ due to reduction of the relative contribution of the
self-phase modulation in comparison with the self-amplitude
modulation.

\begin{figure}
   \begin{center}
   \begin{tabular}{c}
   \includegraphics[height=6cm]{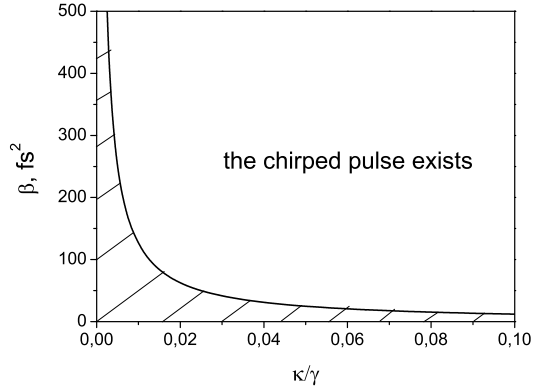}
   \end{tabular}
   \end{center}
   \caption[fig1]
   { \label{fig1}
Region of the chirped pulse existence for the model
(\ref{eq1},\ref{eq3},\ref{eq4}) with $\zeta=$0 and $\alpha=$2.5
fs$^2$.}
   \end{figure}

Fig. \ref{fig1} (unshaded area) shows the region, where the chirped
pulse exists for the model (\ref{eq1},\ref{eq3},\ref{eq4}) with
$\zeta=$0. One can see, that there exists some minimum (threshold)
positive GDD providing the chirped pulse generation (like the case
without the dynamic gain saturation). The threshold GDD decreases
with the self-amplitude modulation growth (the $\kappa$-growth).
Also, one can see that the simplest analytical model does not
predict the pulse destabilization with the GDD growth (i.e., the
stability region confinement), as it takes a place in the experiment
\cite{confined}.

\section{NUMERICAL THEORY OF CPO WITH THE DYNAMIC GAIN SATURATION}
\label{s3}

It is clear, that the dynamic gain saturation in the form of Eq.
(\ref{eq4}) as well as the assumption of $\zeta=$0 do not allow
taking into account the substantial features of CPO dynamics. Hence,
the numerical simulations based on Eq. (\ref{eq1}) and

\begin{equation}\label{numer}
\frac{{\partial g\left( t \right)}} {{\partial t}} = \frac{{P_p }}
{{S_p }}\frac{{\sigma _a }} {{h\nu _a }}\left( {g_{\max }  - g\left(
t \right)} \right) - \frac{{P\left( t \right)}} {{S E_s }}g\left( t
\right) - \frac{{g\left( t \right)}} {{T_r }}
\end{equation}

\noindent are required. The last equation takes into account the
gain dynamics in a quasi-two-level (four-level, in fact) active
medium \cite{gain}. $P_p$ is the absorbed pump power, $S_p$ is the
pump beam area inside an active medium, $\sigma_a$ is the absorption
cross section of active medium, $\nu_a$ is the pump frequency,
$g_{max}$ is the maximum gain, $T_r$ is the gain relaxation time.

The numerical challenge is that $\sqrt{\alpha} \ll T_{cav}$ and the
integro-differential system (\ref{eq1},\ref{numer}) has to be solved
on the grid containing $2^{21}$ points for $T_{cav} \approx$20 ns.
The propagation step $\Delta z$ in the simulation equals to $1/50$
of the cavity round-trip length.

\begin{figure}
   \begin{center}
   \begin{tabular}{c}
   \includegraphics[height=7cm]{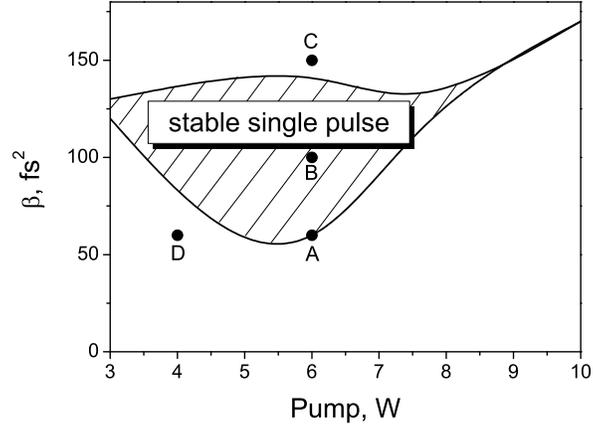}
   \end{tabular}
   \end{center}
   \caption[fig3]
   { \label{f3}
Region of the chirped pulse stability (shaded area) for the
numerical model (\ref{eq1},\ref{numer}) with $\zeta=$0.6$\gamma$,
$\kappa=$0.04$\gamma$, $g_{max}=$0.33, $\l=$0.22, $S_p=$100
$\mu$m$^2$, $S=$130 $\mu$m$^2$, $T_r=$3.5 $\mu$s. Double transit
through the Ti:sapphire crystal per one round-trip ($T_{cav}=$21 ns)
was considered.}
   \end{figure}

The main result of numerical analysis is shown in Fig. \ref{f3}.
Figure presents the stability region on the ``absorbed pump power --
GDD'' plane. One can see, that the stability region is confined:
there are some minimum and maximum GDDs as well as minimum and
maximum pump powers providing the stable chirped pulse generation.
Both minimum and maximum stabilizing GDDs depend on the pump
non-monotonically. That is there exists some optimal pump power
($\approx$5.5 W in our case) providing the pulse stability within
the broadest range of GDDs. When the pump excesses some maximum
value ($\approx$10 W in our case), the pulse cannot be stabilized by
only dispersion adjustment. It should be noted, that the existence
of both maximum GDD and pump, which cannot be exceeded without the
pulse destabilization, is not explicable in the framework of the
previous models. An appearance of these stability limits can be
interpreted as a result of the gain dynamics in a high-energy CPO.

\begin{figure}
   \begin{center}
   \begin{tabular}{c}
   \includegraphics[height=7cm]{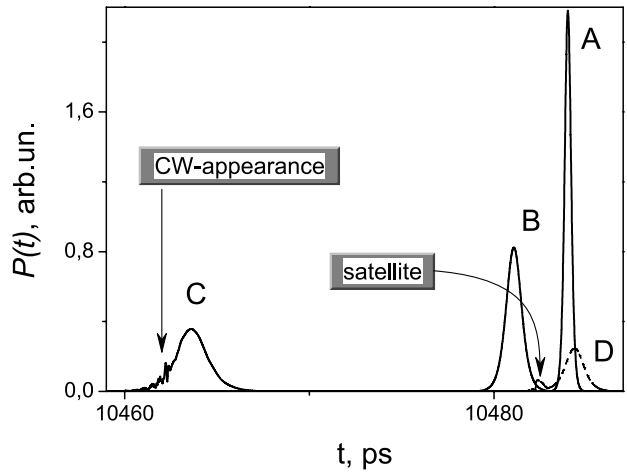}
   \end{tabular}
   \end{center}
   \caption[fig4]
   { \label{f4}
$P(t)$-profiles corresponding to the parameters specified by the
points $A$, $B$, $C$ and $D$ in Fig. \ref{f3}.}
   \end{figure}

\begin{figure}
   \begin{center}
   \begin{tabular}{c}
   \includegraphics[height=6cm]{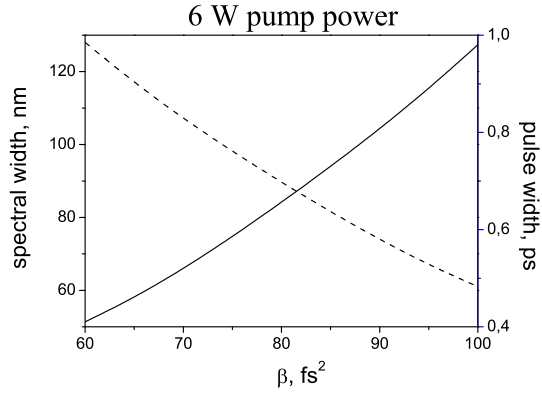}
   \end{tabular}
   \end{center}
   \caption[fig5]
   { \label{f5}
Dependencies of the pulse width (solid curve) and the spectrum width
(dasher curve) on GDD for $P_p=$6 W.}
   \end{figure}

\begin{figure}
   \begin{center}
   \begin{tabular}{c}
   \includegraphics[height=6cm]{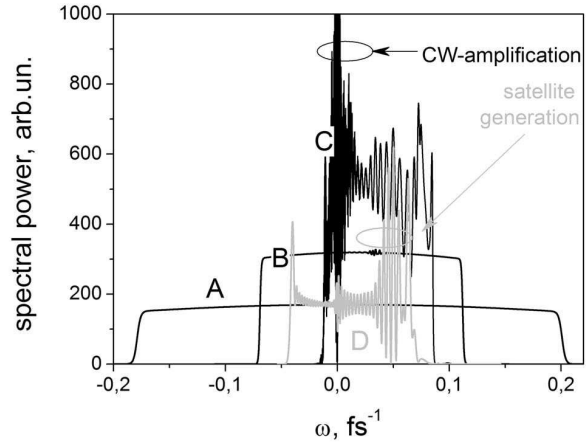}
   \end{tabular}
   \end{center}
   \caption[fig6]
   { \label{f6}
Spectral profiles corresponding to the parameters specified by the
points $A$, $B$, $C$ and $D$ in Fig. \ref{f3}.}
   \end{figure}

Fig. \ref{f4} illustrates the power profiles of both stable ($A$ and
$B$ corresponding to the points in Fig. \ref{f3}) and unstable ($C$
and $D$) pulses. Within the limits of the stability region, the GDD
growth broadens the pulse (transition from $A$ to $B$ in Fig.
\ref{f4}, see, also, solid curve in Fig. \ref{f5}), and reduces the
peak power. The pulse energy does not change substantially, but the
spectral width decreases (dashed curve in Fig. \ref{f5}, see, also
$A$ and $B$ in Fig. \ref{f6}).

The spectra have a profile, which is typical for the CPO: the
parabolic top with the truncated edges ($A$ and $B$ in Fig.
\ref{f6}). In contrast to a situation with no dynamic gain
saturation, the spectra is frequency shifted. This shift results in
a partial compensation of the time-advance caused by the dynamic
gain saturation. Indeed, the last leads to the primary amplification
of the pulse front. The compensation of this effect in the positive
dispersion domain is possible if the pulse spectrum is blue-shifted.

When the GDD crosses the upper stability border, the ripples at the
pulse front appear ($C$ in Fig. \ref{f4}). Such ripples grow and
destroy the pulse. The spectrum ($C$ in Fig. \ref{f6}) demonstrates,
that such a destabilization corresponds to the CW-amplification in
front of the pulse.

On the other hand, both GDD and pump decrease destabilizes the
pulse, as well ($D$ in Figs. \ref{f3},\ref{f4},\ref{f6}). The
destabilizing mechanism is an appearance of satellite in front of
the pulse (dashed curve in Figs. \ref{f3}). As a result of the
dynamic gain saturation, the energy transfer from the pulse to
satellite destabilizes the CPO.

\section{CONCLUSION}
\label{s4}

To determine the operational range of CPO in terms of dispersion and
energy variations, an effect of the dynamic gain saturation on the
CPO stability has been studied numerically for the first time to our
knowledge. It has been found, that the dynamic gain saturation has a
strong impact on a CPO dynamics in contrast to a low-energy (~10 nJ)
mode-locked solid-state oscillator. First, there is some minimum
positive net-GDD providing the oscillator stabilization. Below this
value the pulse is destroyed by excitation of the satellite before
the pulse. Such a satellite is clearly visible in the experiment and
can co-exist stably with the pulse within some narrow GDD-range.
Second, there is some maximum GDD and its excess fragments the pulse
front owing to the CW-amplification. Third, the highest pulse energy
at a given repetition rate is limited because the maximum and
minimum GDDs stabilizing the pulse merge.

\acknowledgments The work was supported by the Austrian National
Science Fund (Fonds zur F\"{o}rderung der Wissentschaftlichen
Forschung (FWF), Projects No. P20293 and P17973) and by the
Max-Planck Gesellschaft (Institut f\"{u}r Quantenoptik). A.Apolonski
(Department f\"{u}r Physik der Ludwig-Maximilians-Universit\"{a}t
M\"{u}nchen) is gratefully acknowledged for active discussions and
providing the experimental data concerning the high-energy
Ti:sapphire CPOs.

{}

\end{document}